\documentclass[aps,prl,twocolumn,,superscriptaddress,showpacs]{revtex4}
\usepackage{hyperref}
\usepackage{amsmath}
\usepackage{amsfonts}
\usepackage{amssymb}
\usepackage{graphicx}
 
\begin{document}
\title{Electron correlation effects in transport and tunneling spectroscopy of the Si(111)-$7\times 7$ surface}
\author{A.B. Odobescu}
\email[]{arty@cplire.ru} 
\affiliation{Kotel'nikov Institute of Radioengineering and Electronics of RAS, Mokhovaya 11, bld.7, 125009 Moscow, Russia}
\affiliation{Moscow Institute of Physics and Technology, 141700 Dolgoprudny, Russia}

\author{A.A. Maizlakh}
\affiliation{Kotel'nikov Institute of Radioengineering and Electronics of RAS, Mokhovaya 11, bld.7, 125009 Moscow, Russia}
\affiliation{Moscow Institute of Physics and Technology, 141700 Dolgoprudny, Russia}

\author{S.V. Zaitsev-Zotov}
\affiliation{Kotel'nikov Institute of Radioengineering and Electronics of RAS, Mokhovaya 11, bld.7, 125009 Moscow, Russia}
\affiliation{Moscow Institute of Physics and Technology, 141700 Dolgoprudny, Russia}

\date{\today}

\begin{abstract}
Electronic properties of the Si(111)-$7\times 7$ surface are studied using four- and two-probe conductivity measurements and tunneling spectroscopy. We demonstrate that the temperature dependence of the surface conductivity corresponds to the Efros-Shklovskii law at least in $10-100$~K temperature range. The energy gap at the Fermi level observed in tunneling spectroscopy measurements at $T\geq 5$~K vanishes by thermal fluctuations at $T\approx 30$~K, without any sign of the metal-insulator transition. We show that the low-temperature energy gap observed by the tunneling spectroscopy technique is actually the consequence of the Coulomb blockade effect.
\end{abstract}
\maketitle

\section{Introduction}
Electron transport in low-dimensional materials is one of the major topics of condensed matter physics. Reducing dimensionality of physical objects may change qualitatively their properties. The atomic size scale is the natural limit for size reduction. Massless Dirac particles in graphene \cite{graphen}, surface superconductivity \cite{supcond}, various one-dimensional systems with Luttinger liquids \cite{luttinger} are well known examples. At such scales electron correlation effects often start to dominate.

The self-assembled superstructures on semiconductor surfaces have wide potentialities for experimental study of the electron transport in one- and two-dimensional systems. The scanning tunneling spectroscopy (STS) is often used to identify the electronic structure of the ground state of these surfaces with atomic resolution. But the electron correlations can induce appreciable distortions in the STS spectra \cite{levitov}. The effect is enhanced on the low conductive system with sheet conductivity much less than minimum metallic conductivity $1/R_h =e^2/h = 39 \mu S$.

A striking example is the Si(111)-$7\times7$ surface, one of the most comprehensively studied structures. On the one hand, the surface has a metallic band structure revealed by the low temperature angle-resolved photoemission spectroscopy (ARPES) \cite{losio}.  On the other hand, a hard energy gap at the Fermi level ($E_F$) is observed at low temperature with STS \cite{modesti, odobescu}. The experimental data from nuclear-magnetic-resonance \cite{nmr}, high resolution electron-energy-loss spectroscopy \cite{persson}, high resolution photoemission \cite{barke}, and transport measurements with micro four-point probe (MFPP) technique \cite{hasegawa} also hint to a gap between $0$ eV  to $0.2$ eV at low temperatures. Possible reasons for the gap opening  are suggested to be formation of the charge-density waves or Mott-Hubbard insulator ground state. Similar discrepancy between metallic surface band structure revealed by ARPES \cite{uhrberg} and nonmetallic electron transport behavior \cite{hirahara}, a hard gap singularity at $E_F$ by STS at low temperatures \cite{modesti_sn} was also observed for the Si(111)-$\sqrt{3}\times\sqrt{3}$-Sn surface.

In this paper we report the results of detailed study of both the temperature evolution of the tunneling density of states (TDOS) and conductivity of the Si(111)-$7\times7$ surface in a wide-temperature range. No evidence of any phase transition for this surface down to helium temperatures is observed. Instead, we found that the temperature variation of the surface conductivity follows the Efros-Shklovskii law at least in $10-100$~K temperature range, the surface conductivity value being changed over eight orders of magnitude. The shape of Coulomb gap for 2D, which follows from the observed Efros-Shklovskii conduction, is in obvious contradiction with the energy gap in the local density of states found earlier \cite{modesti, odobescu}. We argue that the hard-like energy gap observed in STS does not correspond to the true local density of states, but is actually the Coulomb gap modified by the local charging effect in the surface. We consider that similar disagreement between nonmetallic conductivity and the gap in STS spectra at low temperatures are common for most low-conductive surface reconstruction with `metallic' surface band (Sn, Ag, K on Si(111) and Ge(111) and others)  and could be explained by localization and Coulomb blockade effect. 

\section{Experimental}
We studied n-type Si(111) samples with $\rho = 0.1$ $\Omega\cdot$cm and $\rho = 1$ $\Omega\cdot$cm for transport measurements. The clean Si(111)-$7\times 7$ surface was prepared by direct current degassing at $T \approx 900$~$^o$C for 12 hours followed by flashing  up to 1250$^o$C for 10 seconds and slow computer-controlled cooling down. The quality of the Si(111)-$7\times 7$ surface was controlled by STM. 
Conduction measurements were performed {\em in situ} in the  home-built  four-probe system \cite{pribandexp}, mounted on a cold finger of UHV cryogenerator SHI-APD Cryogenics. The system operates from room temperature to the 35 K. The probes were made from optic fiber wires ($d=20 \mu$m) covered with Ta conduction film. The distance between the probes was $100 \pm 20 \mu$m. Four conducting probes assembled on the special holder in the linear geometry were approached to the surface by piezodrivers simultaneously until electric contact to the surface was obtained for all the probes. The current {\it I} flows between the outer pairs of probes and the voltage drop $\Delta V$ was measured between the inner pair of probes. The conductivity 
$$\sigma_{4p} = \frac{\ln 2}{\pi }\frac{I}{\Delta V}$$
was measured from the slope of the linear current-voltage curves ({\it I-V} curves) collected during slow temperature sweeps, which takes about 6 hours in one direction. Accuracy of the temperature measurements was verified by a calibrated thermometer mounted onto the sample holder and placed instead of a silicon sample.
The two-probe measurements where performed separately using Omicron LT SPM with modified standard tip holder, equipped with two $50{\ }\mu$m platinum wires. This method let us to extend the range of the studied conductivity down to $10^{14} \Omega$, which is hardly available for the four-probe study. The distance between contacts {\it l} was 1 mm. The {\it I-V} curves were measured at fixed sample temperatures.  
The conductivity is
$$\sigma_{2p} = \frac{I}{\pi V}\ln(\frac{l}{d})$$
where $d$ is a characteristic size of the contact area. This equation accounts for the geometrical spreading resistance and implies absence of an energy barrier between the platinum wires and the surface. We take $\ln(l/d)\approx 10$ which corresponds to $d=1-1000$~nm within 30\% accuracy.

\begin{figure}[h]
 \begin{tabular}{c}
	\resizebox{70mm}{!}{\includegraphics{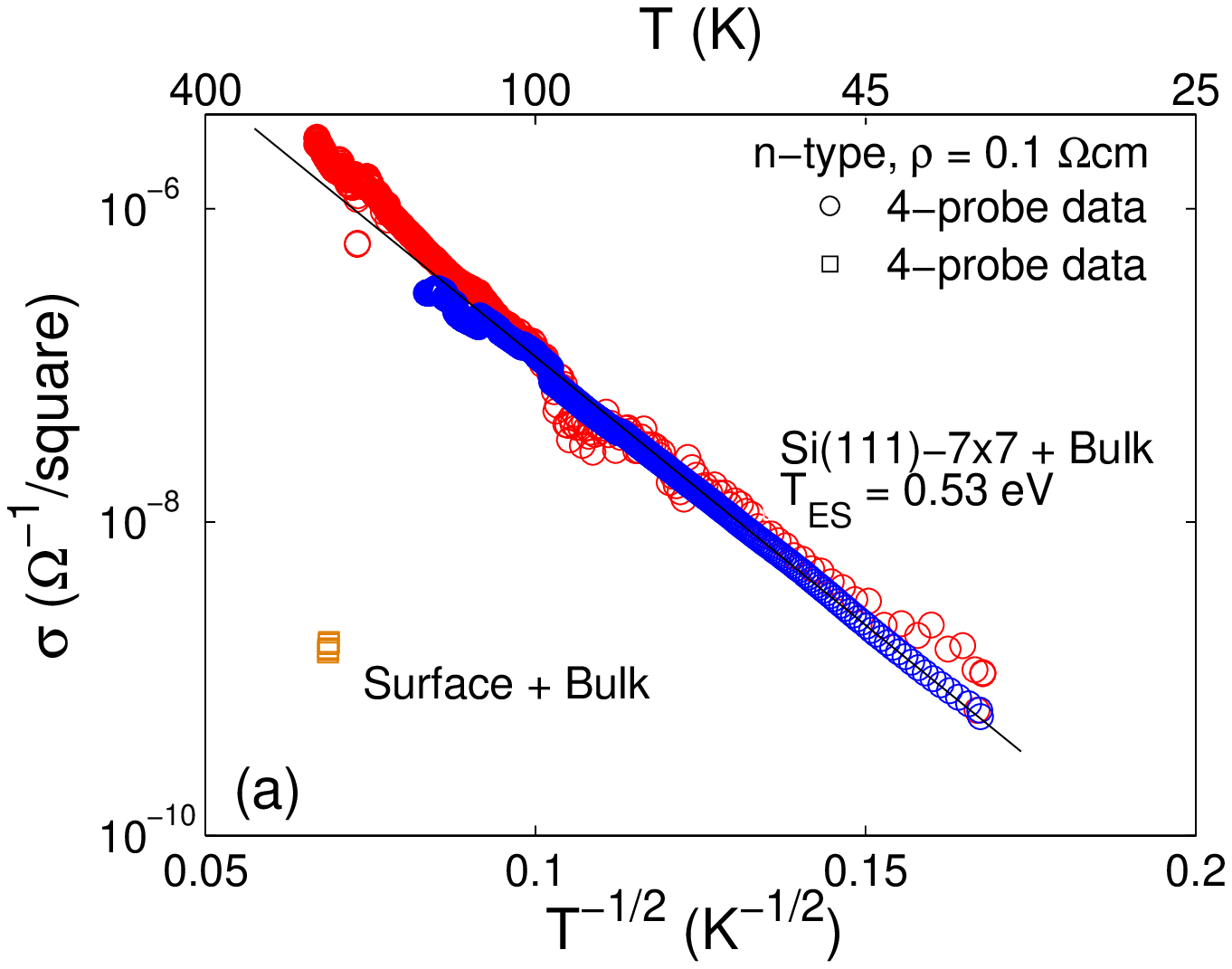}} \\
	\resizebox{70mm}{!}{\includegraphics{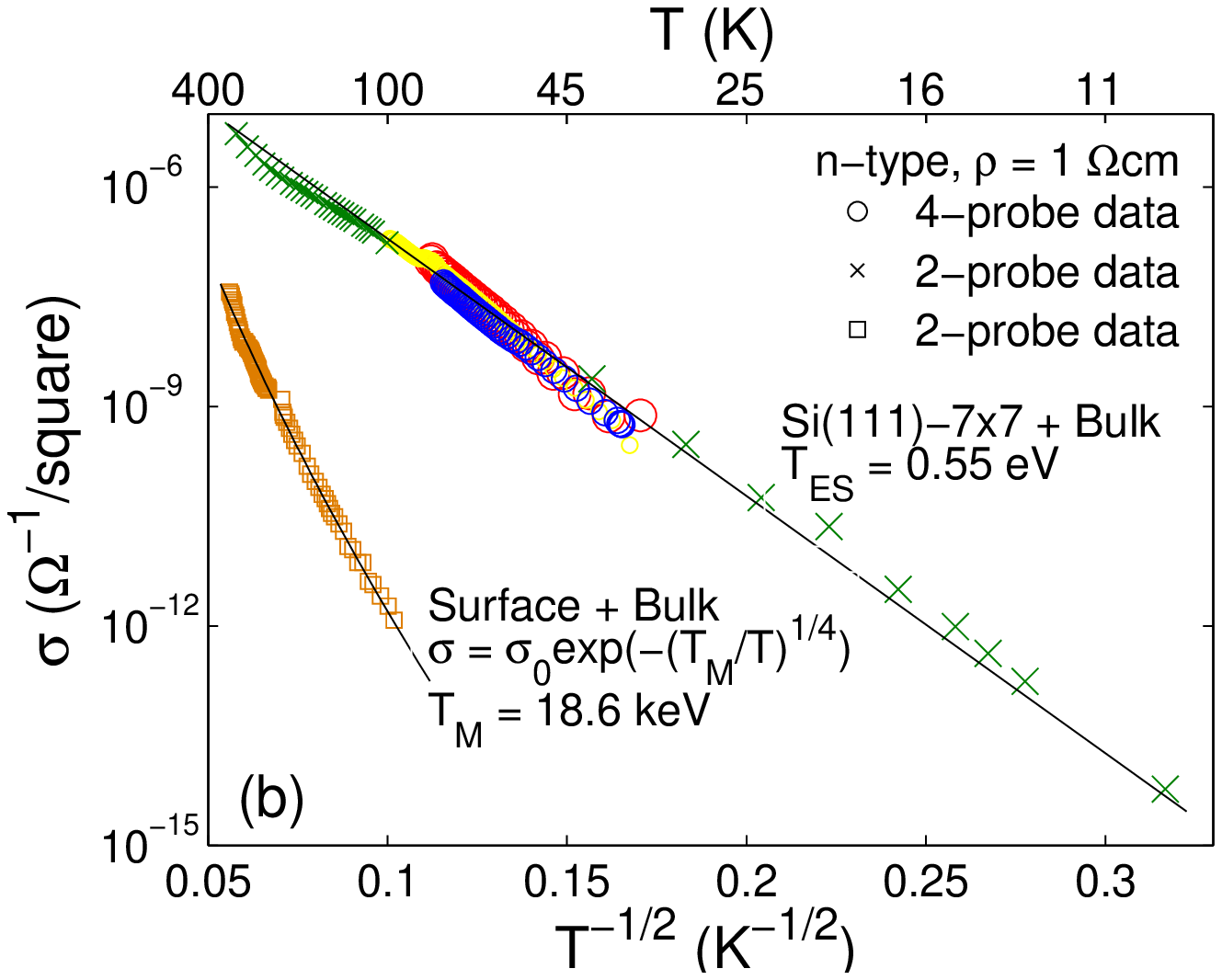}}
 \end{tabular}
\caption{Temperature variation of four- and two-probe conductivity obtained for freshly prepared reconstructed Si(111)-$7\times7$ surfaces (circles and crosses) and for degraded surface (squares, see text for details). (a) four-probe data for n-type Si, $\rho = 0.1$~$\Omega\cdot$cm, $T = 35-250$~K; b) four-probe and two-probe data for n-type Si, $\rho = 1$~$\Omega\cdot$cm, $T = 10-300$~K. In both cases the conductivity of the Si(111)-$7\times7$ surface corresponds to the Efros-Shklovskii law (\ref{eq:es}) with $T_{ES} \sim 0.5$~eV.}
\label{cond_compare}
\end{figure}

The measured conductivity contains the contributions of the conduction through the topmost layer of the Si(111)-$7\times7$ surface, as well as through the bulk and the sub-surface layer \cite{hasegawa, wells, hasegawa2}. The sub-surface layer is known to be a depletion layer due to the Fermi-level pinning by surface state. Since the distance between the probes is much larger than the depletion layer thickness ($<10\mu m$) the total conductance is the sum of the conductance of the Si(111)-$7\times7$ surface plus conductance of the bulk in series with sub-surface layer. In order to estimate the bulk and sub-surface layer contribution we performed control experiments by measuring changes in conductivity during both momentary and continuous surface degradation. A short-time vacuum degradation up to $p=10^{-8}$~Torr at $T = 200$~K reduces the four-probe measured conductivity by more than 3 orders of magnitude (Fig.\ref{cond_compare}~(a)). The two-probe temperature dependent conductivity value as a function of time was found to drift smoothly downwards during a long-time exposition of the Si(111)-$7\times7$ surface under UHV conditions (up to 6 days at $p\approx 10^{-10}$~Torr). The best fit of the data for degraded surface is given by Mott law for 3D system, $\sigma\propto\exp [-(T_M/T)^{1/4}]$ (squares in Fig.\ref{cond_compare}~(b)). Thus the observed contribution of the bulk conduction in the temperature range studied is negligible, below 1\%, and a conduction of the Si(111)-$7\times7$ surface is dominating. 

The tunneling spectroscopy measurements were performed in UHV LT STM  Omicron. We studied {\it n-} and {\it p}-type Si with $\rho = 1$~$\Omega\cdot$cm. Such crystals are insulating in the low-temperature region and cannot be studied by the standart methods. Therefore, for STS study we used the external illumination to produce necessary bulk conduction (see details in \cite{odobescu}). All STS measurements were performed with platinum STM tips. The {\it I-V} curves were collected at fixed temperature over the central adatom of the Si(111)-$7\times7$ unit cell and averaged over series of measurements consisting of tens individual cycles each of 1000 points back and forth. Differential conduction was calculated numerically from the averaged {\it I-V} curves.

\section{Results}

Figure~\ref{cond_compare} shows the temperature dependencies of the Si(111)-$7\times 7$ surface conductivity measured on n-type samples with different room-temperature resistivity. The conductivity variation exceeds eight orders of magnitude in the temperature range $10-250$~K. Our results indicates that the Efros-Shklovskii law 
 \begin{equation}
  \sigma = \sigma _{0}\exp\left( - \left[ \frac{T_{ES}}{T}\right]^{\frac{1}{2}}\right)
  \label{eq:es}
 \end{equation}
fits conductivity of the Si(111)-$7\times 7$ surface much better than any other law, such as $\sigma \propto \exp(-[T_0/T]^{1/3})$ corresponding to the Mott law conduction in 2D or a simple activation law as it was proposed in the previous publication \cite{hasegawa}, where the temperature dependence was studied at $T > 100$~K and conductivity variation slightly exceeds one order of magnitude. 

\begin{figure}[!h]
  \begin{tabular}{cc} 
  	\resizebox{43mm}{!}{\includegraphics{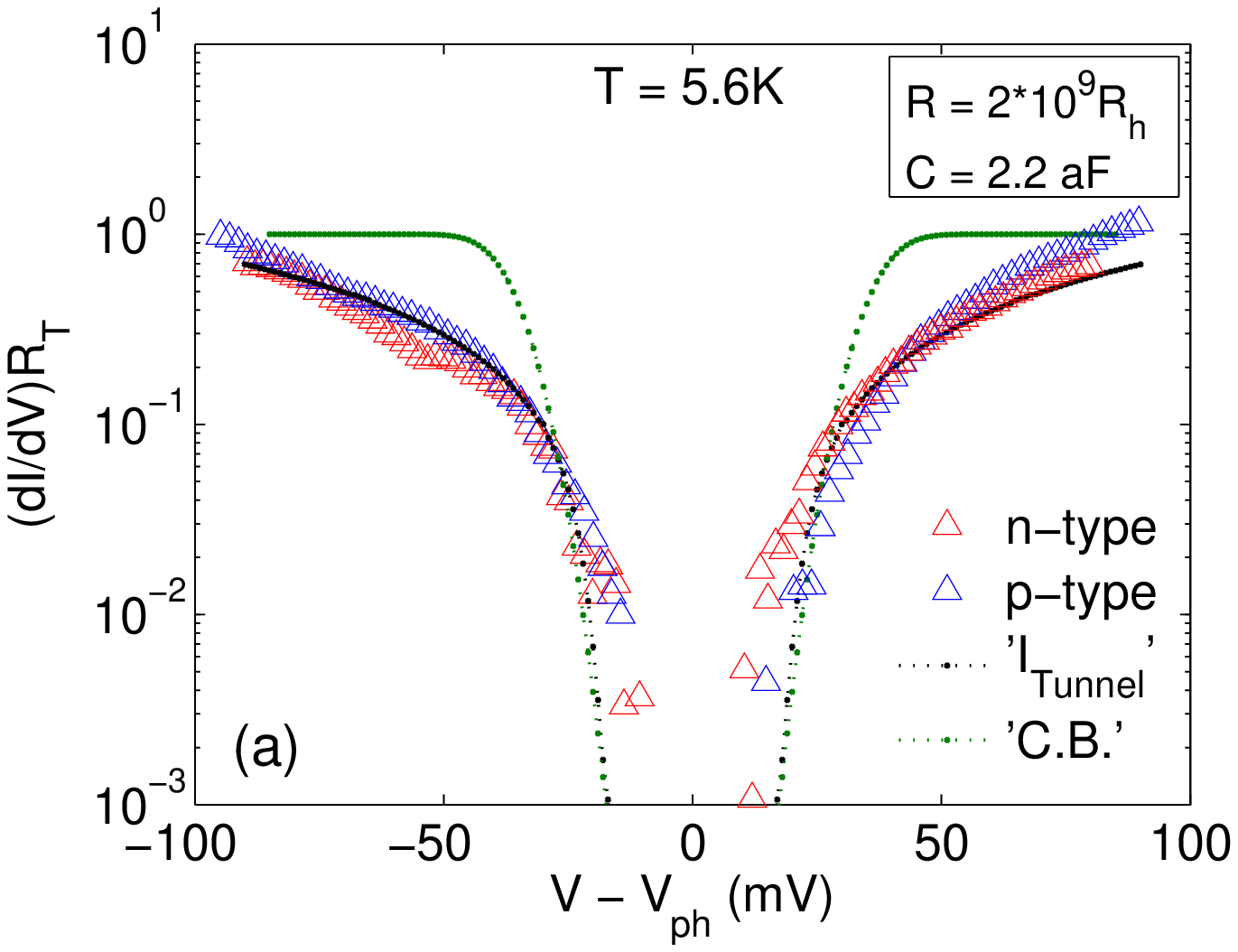}} & \resizebox{43mm}{!}{\includegraphics{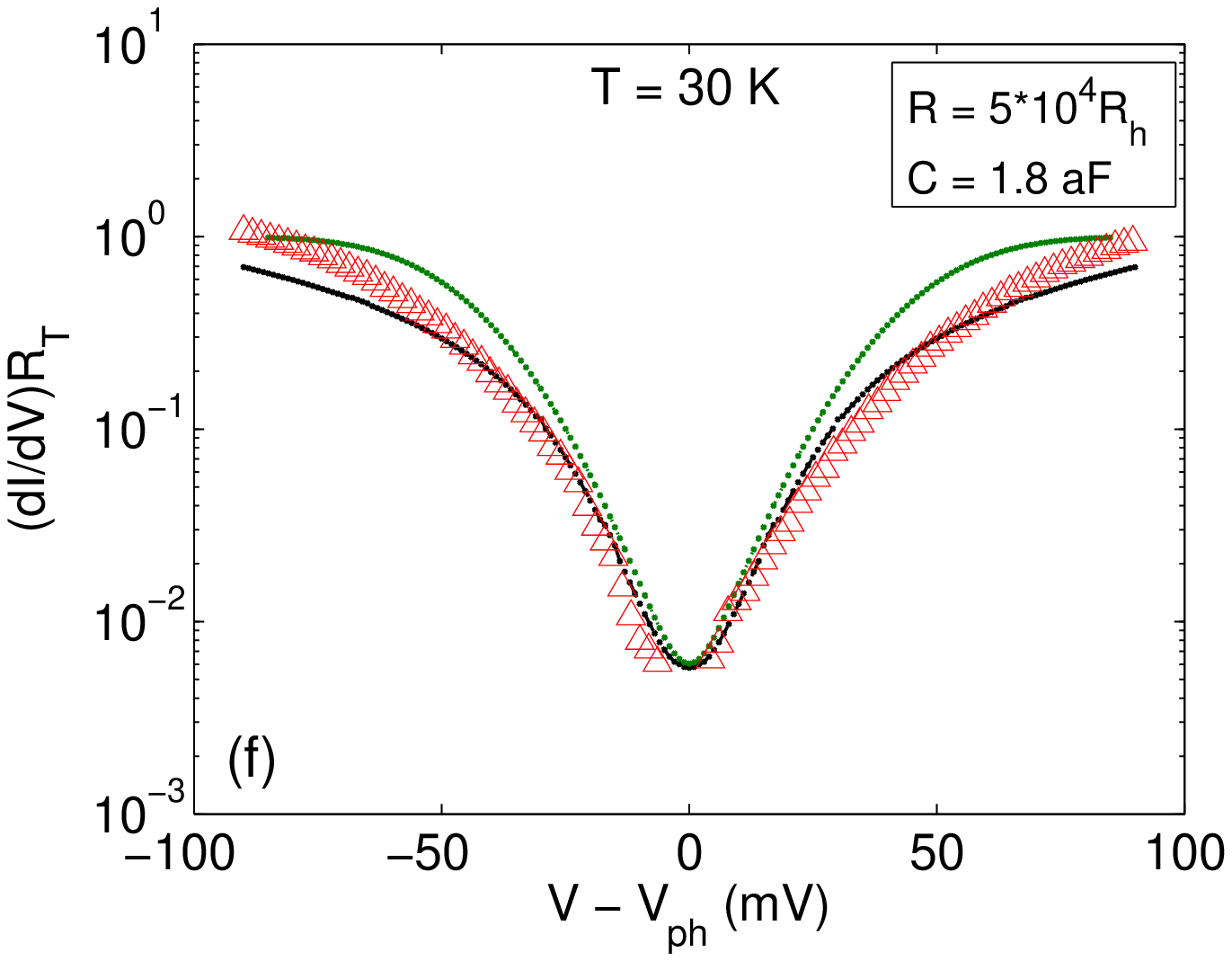}} \\
  	 \resizebox{43mm}{!}{\includegraphics{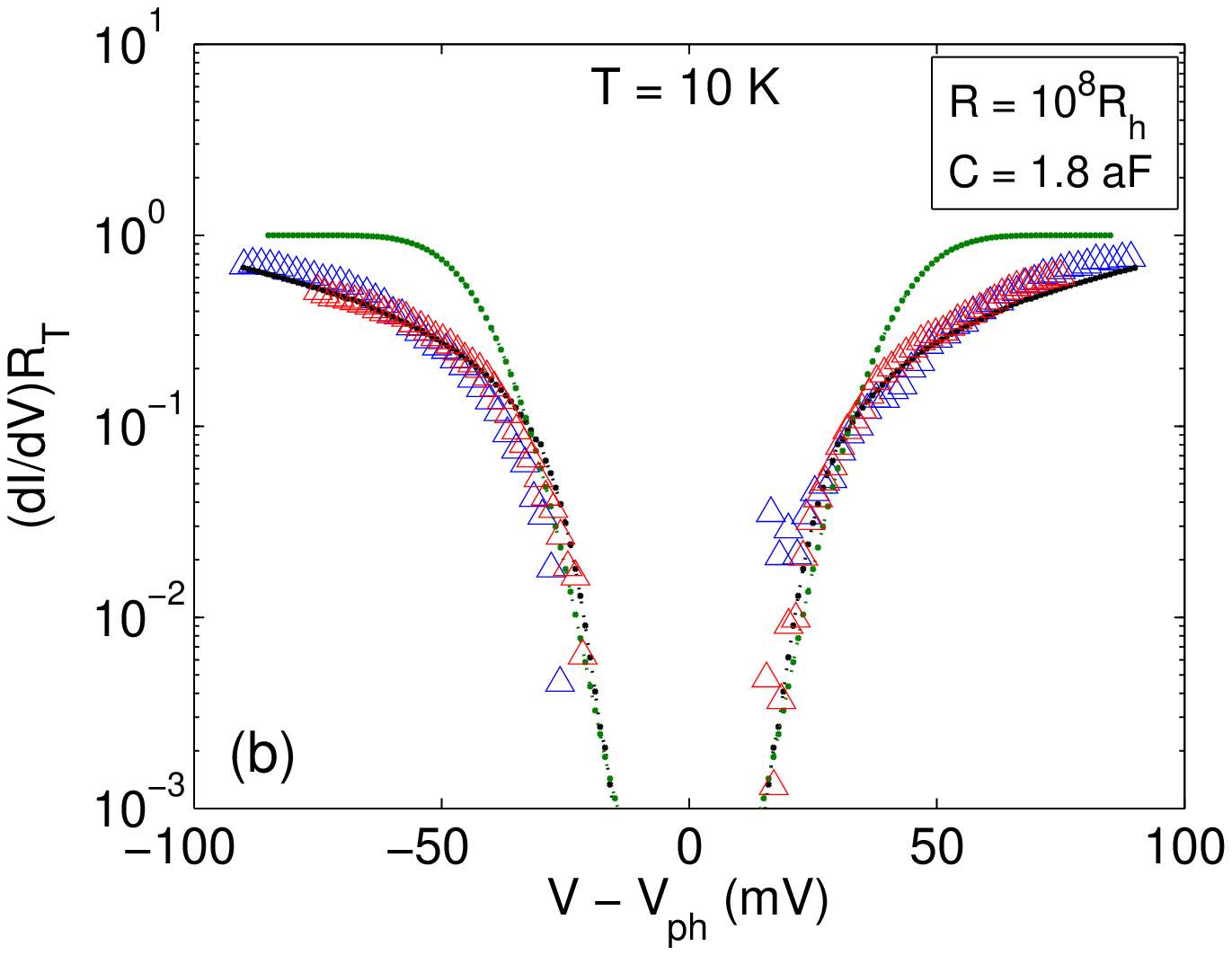}} & \resizebox{43mm}{!}{\includegraphics{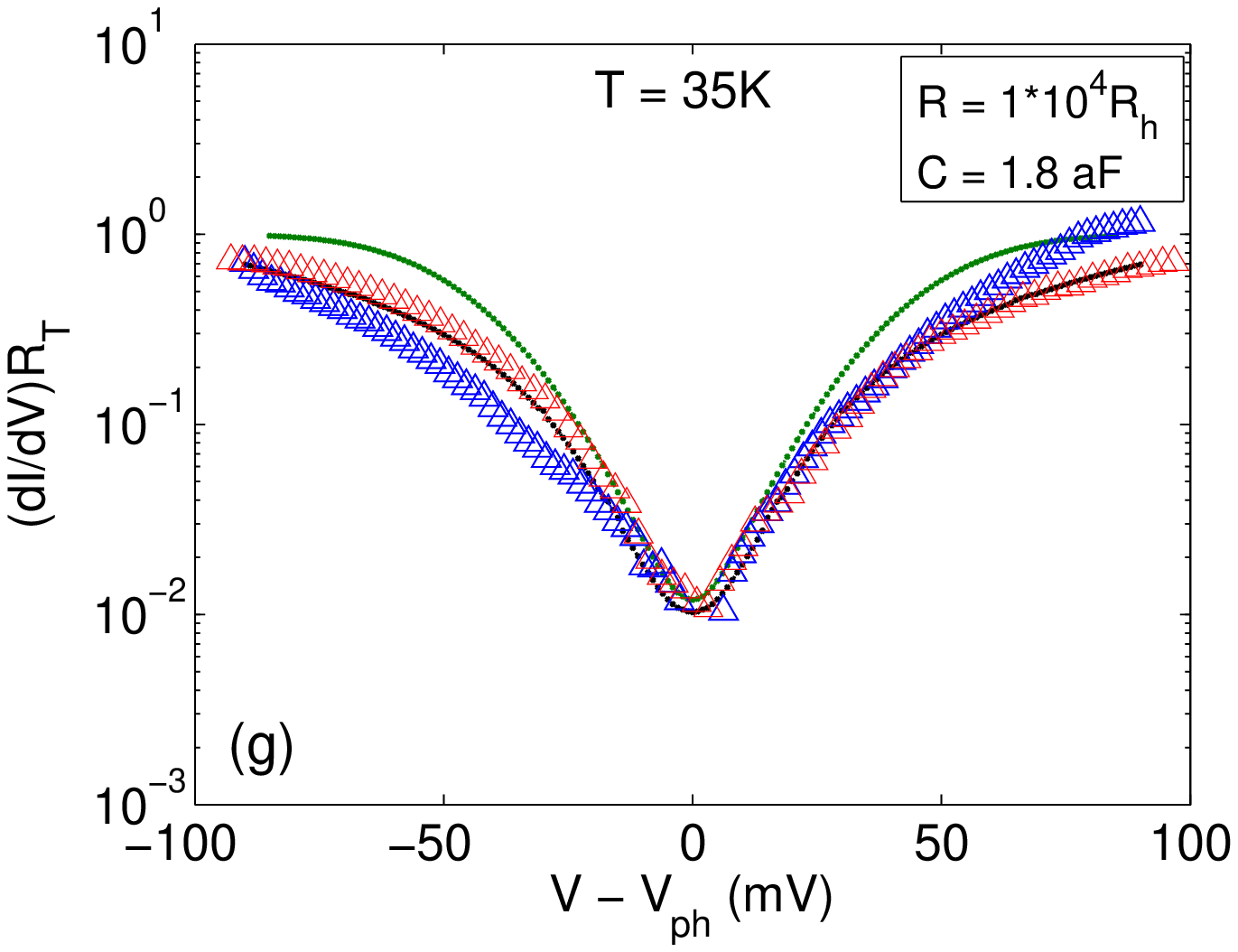}} \\
  	 \resizebox{43mm}{!}{\includegraphics{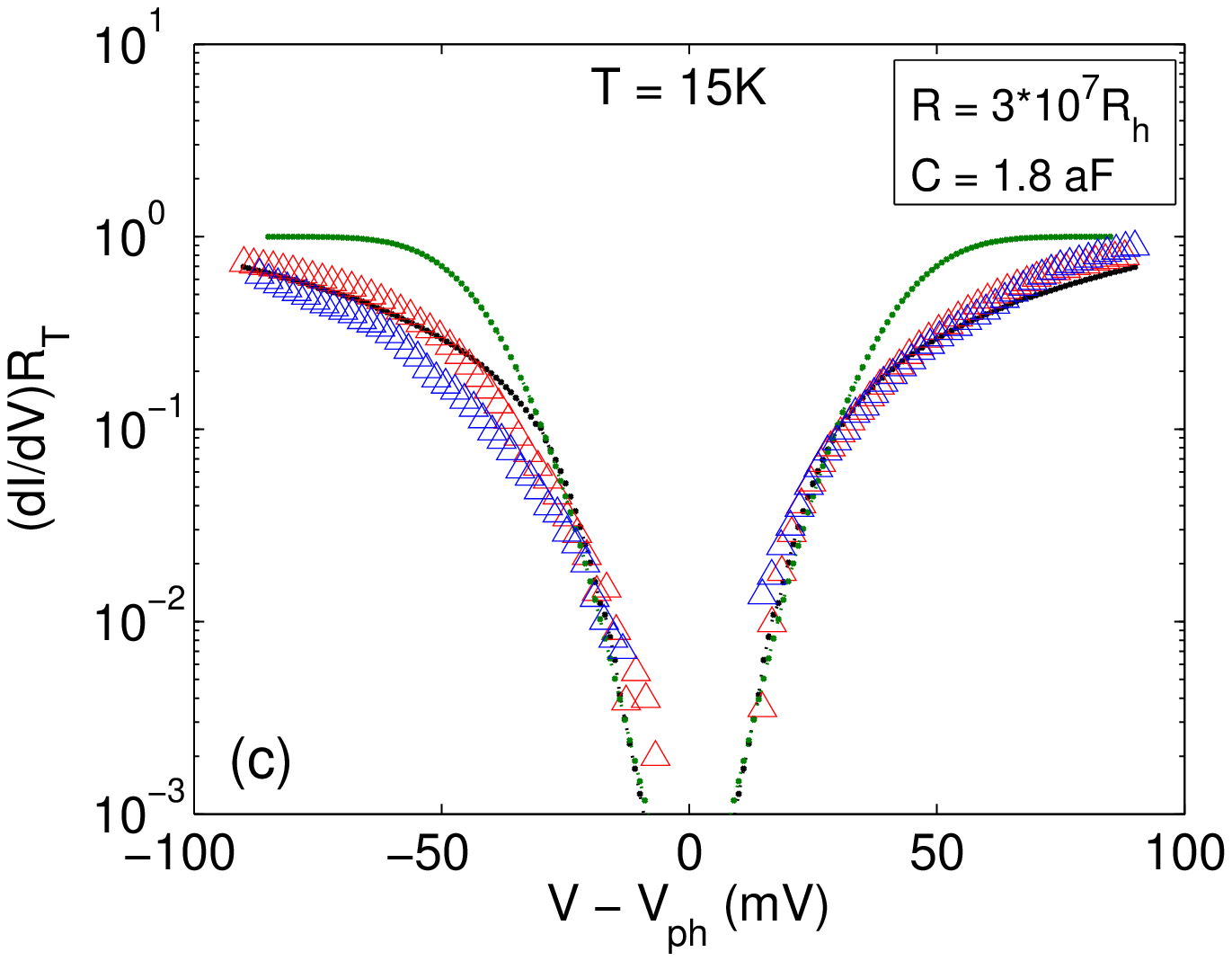}} & \resizebox{43mm}{!}{\includegraphics{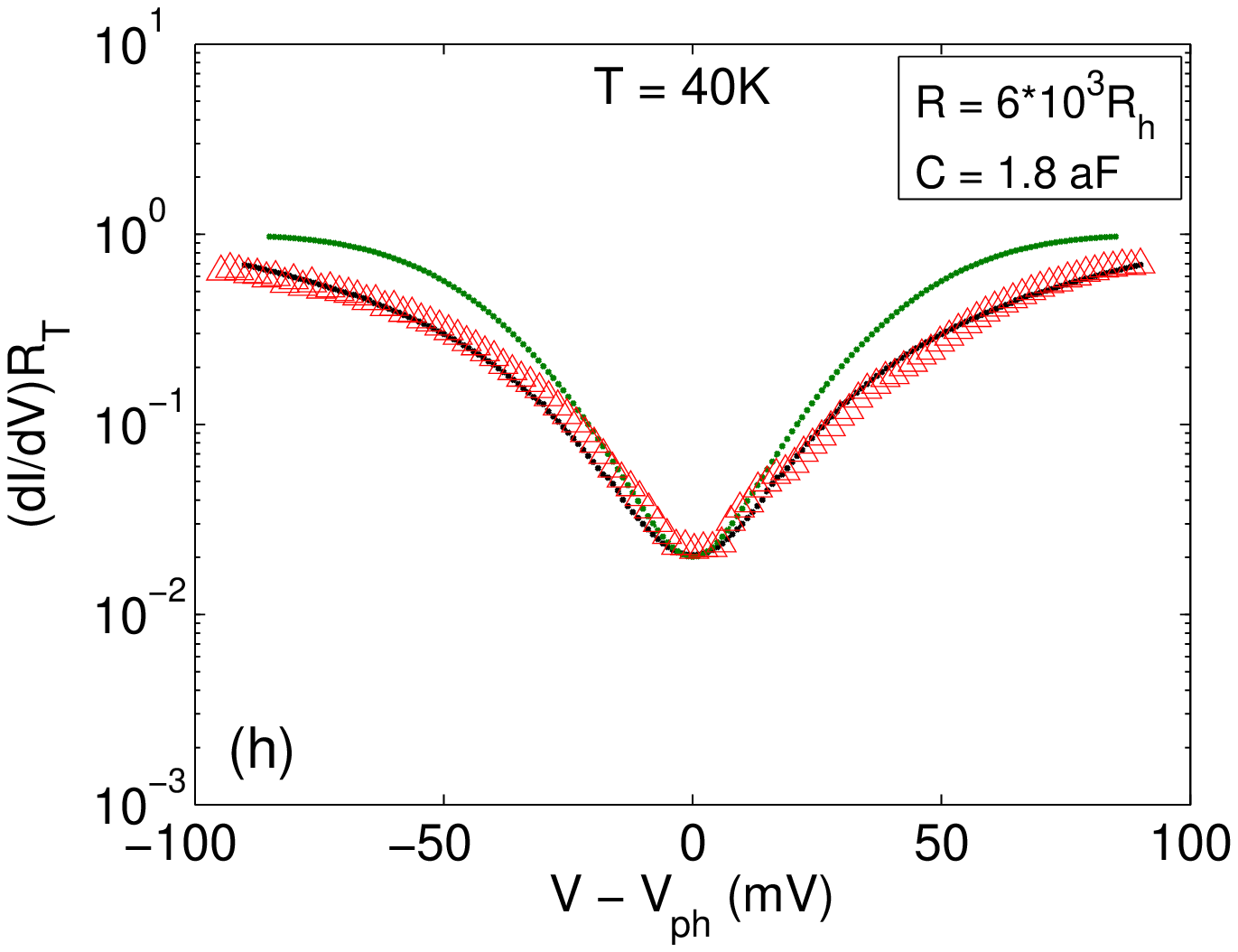}} \\
  	\resizebox{43mm}{!}{\includegraphics{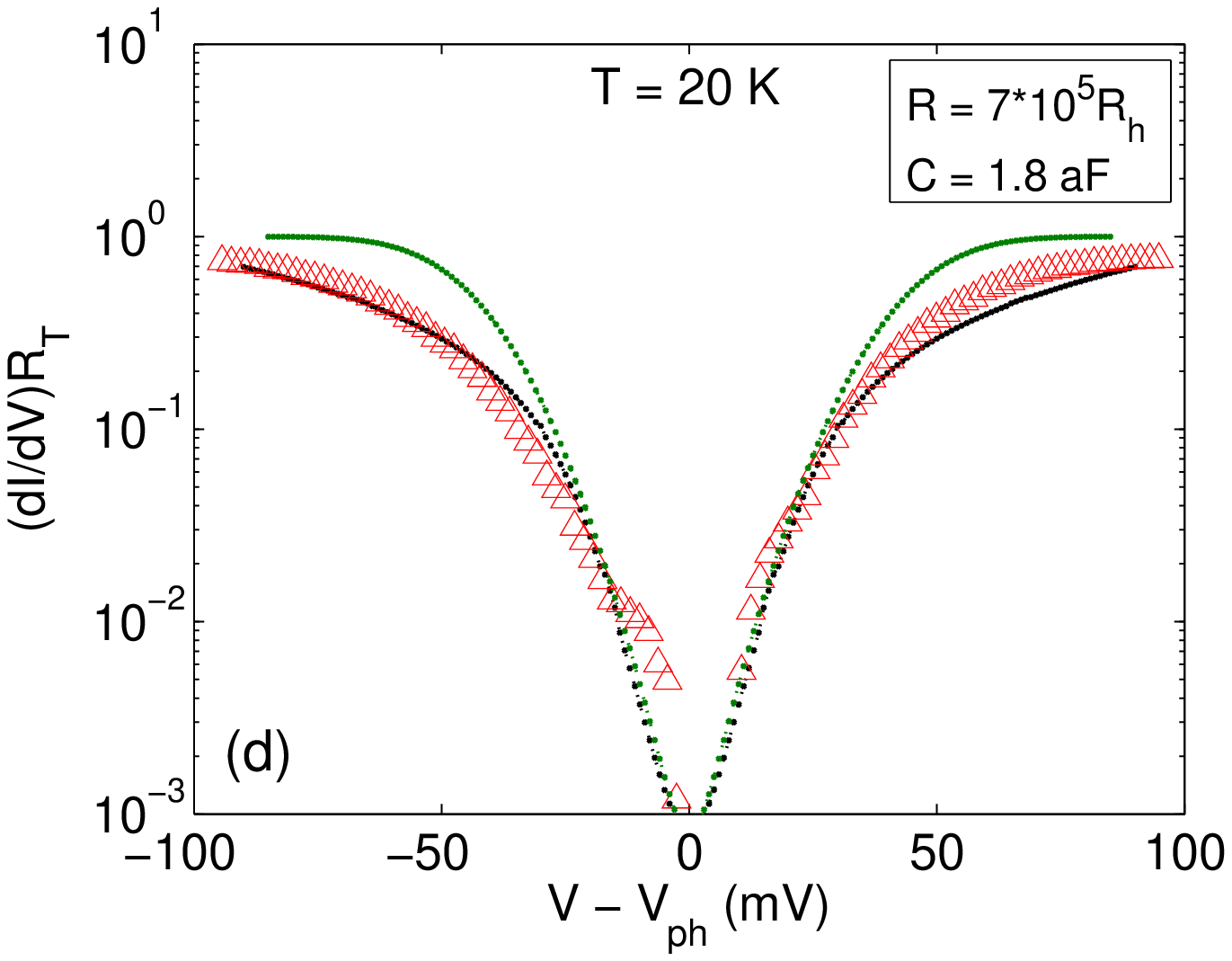}} & \resizebox{43mm}{!}{\includegraphics{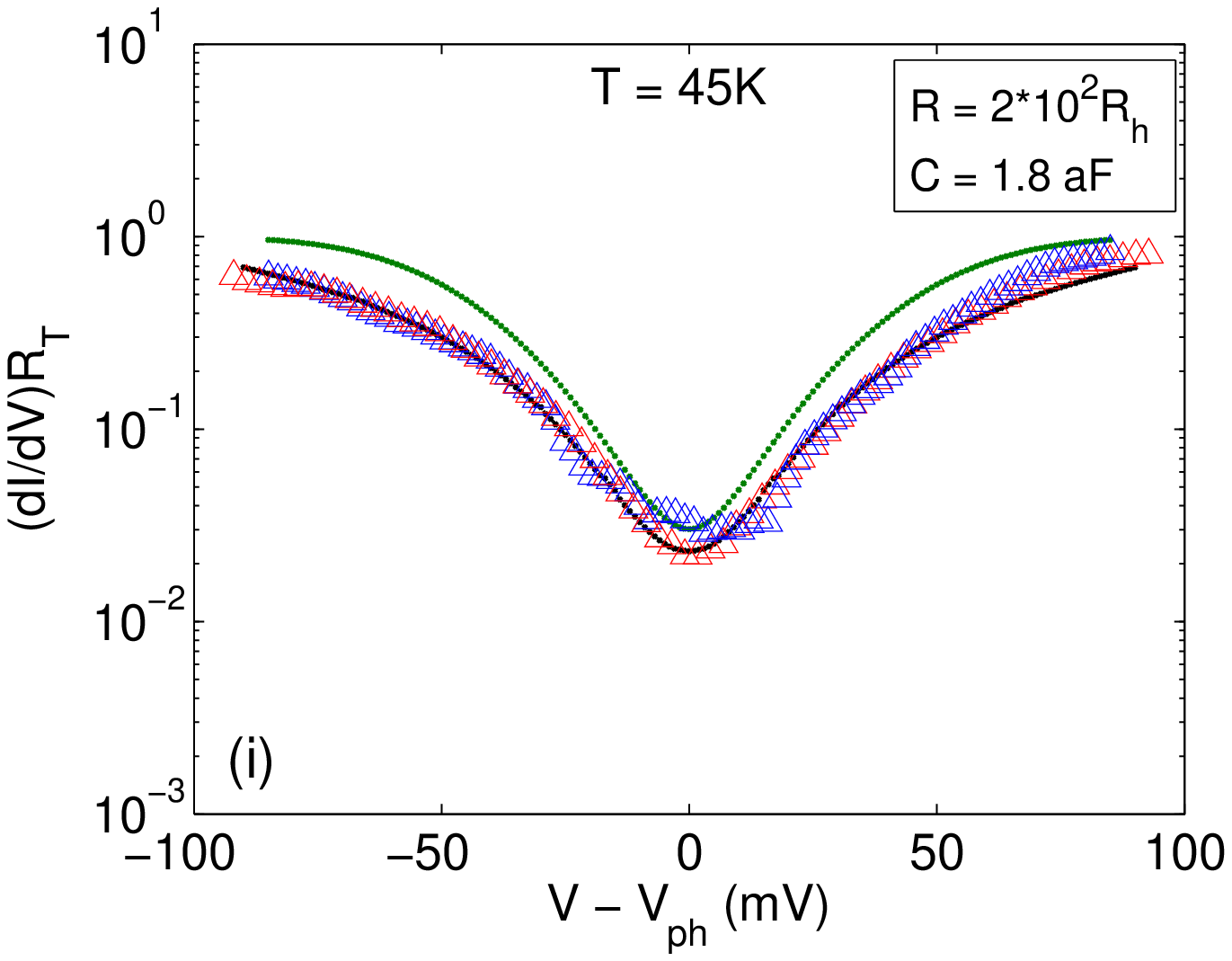}} \\
  	\resizebox{43mm}{!}{\includegraphics{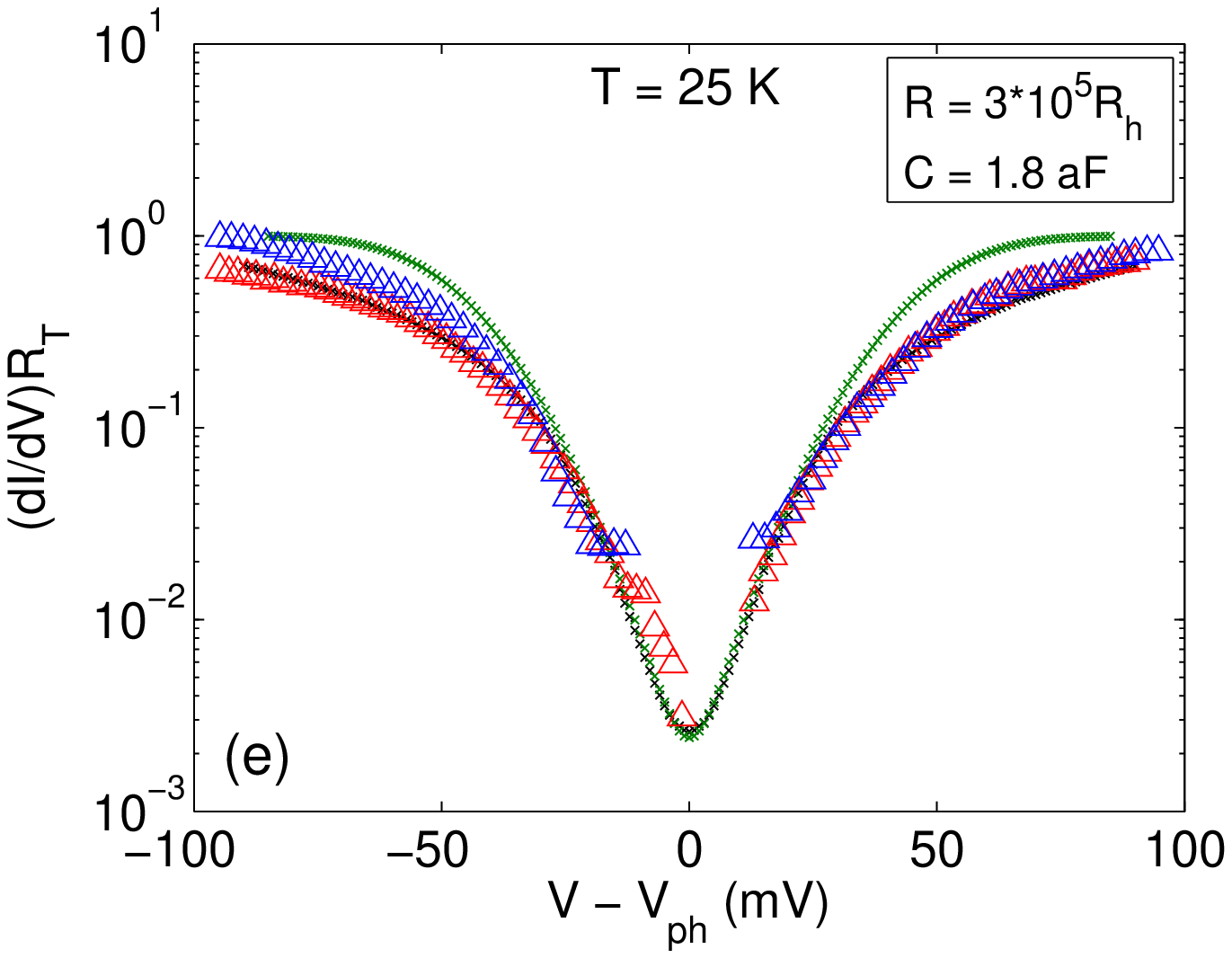}} & \resizebox{43mm}{!}{\includegraphics{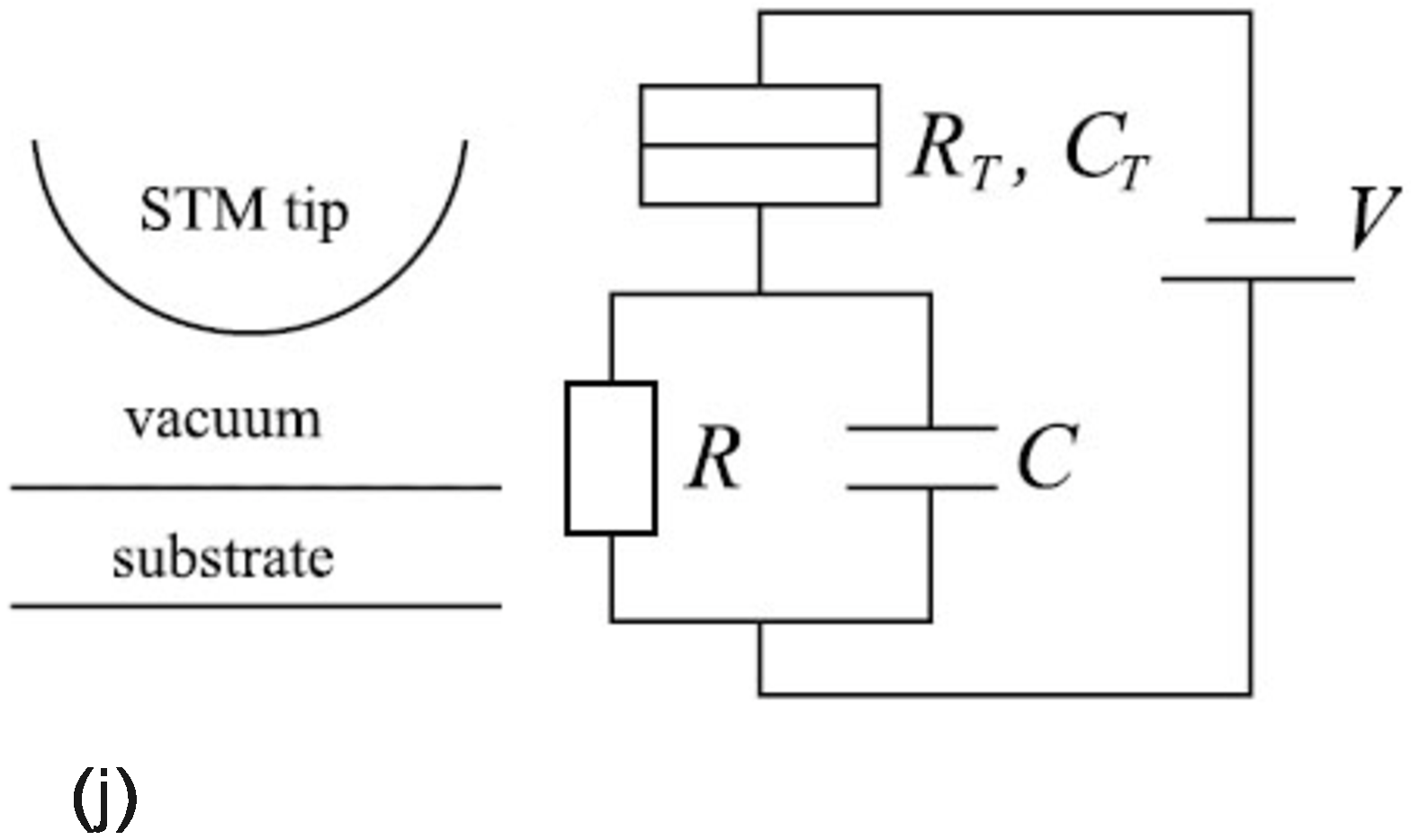}} 
  \end{tabular}
\caption{(a-i)Temperature evolution of the energy gap at the Fermi level. Fragments of $dI/dV$ curves of the illuminated \textit{n}-type (red markers) and \textit{p}-type (blue markers) Si(111)-$7\times 7$ surface, with  $\rho = 1$~$\Omega \cdot$cm in the gap region shifted by the surface photovoltage $V_{ph}$  (+0.6~eV for \textit{p}-type and -0.4~eV for \textit{n}-type samples at $T = 5$~K \protect\cite{odobescu}). The curves are normalized by ${R_T^{-1}}$, where $R_T = V_0/I_0$ is the resistance of the tunneling junction at the set point $I_0 = 100$ pA, $V_0 = 2$ V. Green dotted line is the theoretically calculated $dI/dV$ characteristic of the tunnel junction coupled to the environment characterized by $R$ and $C$, the fitting parameters are shown in the insets.  Black dotted line is the $dI/dV$ characteristic of the tunneling junction calculated for the model shown on the Fig.~\ref{enDiakT}.  (j) The electrical circuit of the tunneling junction used for Coulomb blockade approximation.}
\label{LDOSdiffT}
\end{figure}

Temperature sets of the tunneling $dI/dV$ spectra near the surface Fermi level measured on both {\it n}- and {\it p}-type Si(111)-$7\times7$ samples with $\rho = 1$~$\Omega\cdot$cm are shown in Fig.~\ref{LDOSdiffT}. The Fermi level is shifted by the surface photovoltage due to external illumination \cite{odobescu}. The low-temperature data obtained at $T = 5.6$~K corresponds to the energy gap $2\Delta = 40 \pm10$ meV reported earlier \cite{odobescu}.  We see that the  gap smears quickly with the temperature increase and nonzero DOS at the Fermi level is getting visible for both for \textit{n-} and \textit{p-}type silicon samples at $T \gtrsim 30$~K.

\section{Discussion}
According to \cite{shklovskii}, in the system described by hopping conduction, the electrons are localized and this is the Coulomb interaction that turns the conductivity into the Efros-Shklovskii conductivity regime. Experimental value of the parameter $T_{ES} = 0.55 \pm 0.05$ eV is approximately the same for all studied crystals for both 4- and 2-probe techniques. The activation energy at $T = 10$ K, $\frac{1}{2}\sqrt{T_{ES}T} \approx 10$ meV, is noticeable lower than $\Delta$. The localization length of electrons is
$$
\xi = \frac{2.8e^2}{\kappa T_{ES}}=1.3 \ nm 
$$ 
where $\kappa=(11.8+1)/2$, because of the vacuum above the Si surface, $11.8$ is the Si dielectric constant. The obtained value is in a good agreement with the model \cite{ortega}, where electrons responsible for the surface transport are localized on adatom dangling bonds  and the electron localization length matches  the distance of a unit cell between two central adatoms, $l \approx 0.6$ nm length.

\begin{figure}
 \begin{tabular}{cc}
	\resizebox{42mm}{!}{\includegraphics{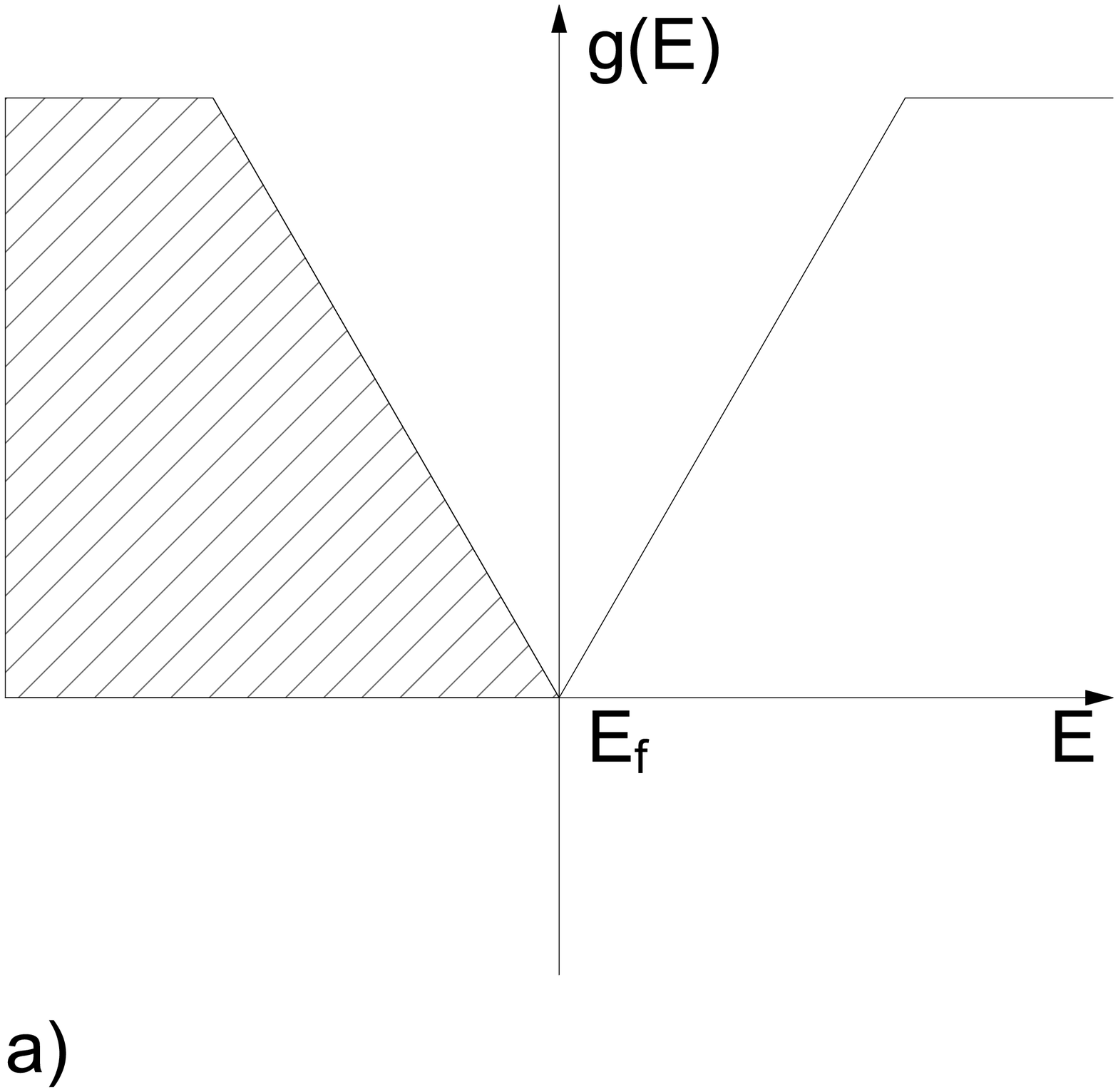}} &
	\resizebox{42mm}{!}{\includegraphics{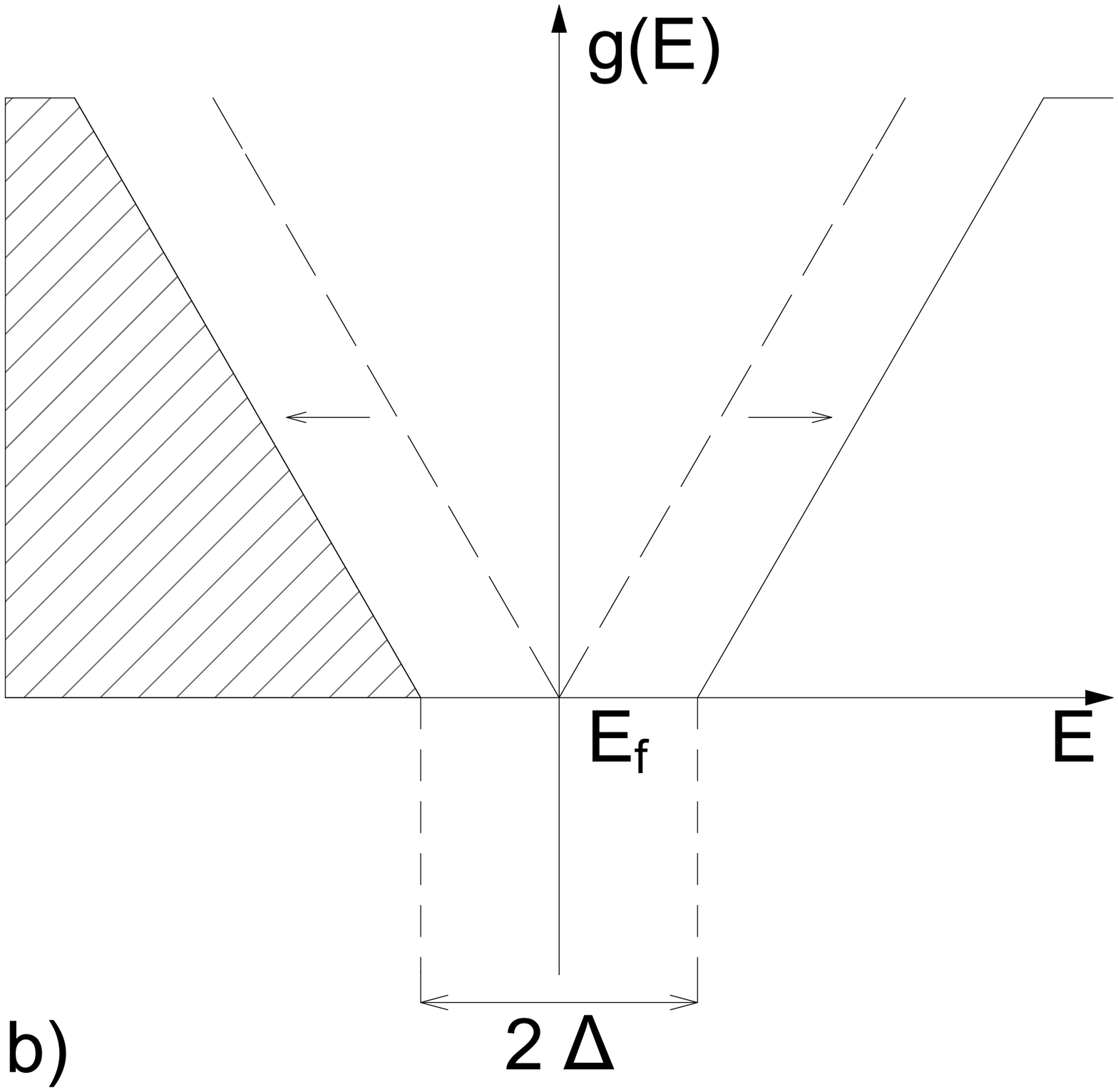}}
 \end{tabular}
\caption{a) The Coulomb gap for 2D system with localized charge carriers (Eq. \ref{eq:DOS}) \cite{shklovskii} b) the proposed modification of the Coulomb gap in the tunneling spectroscopy measurements due to the local environmental Coulomb blockade.}
\label{gapDiagr}
\end{figure}

The tunneling data seems to be in the obvious contradiction with the measured temperature variation of the surface conductivity and band structure obtained by ARPES \cite{losio}. Indeed, the Efros-Sklowskii law (\ref{eq:es}) corresponds to the soft gap in the DOS, $g(E)$, at the Fermi level (Fig.~\ref{gapDiagr}(a)) \cite{shklovskii}
\begin{equation}
g(E) \sim \frac{2}{\pi }\frac{\kappa^2}{e^4}\left|E - E_F \right|,
\label{eq:DOS}
\end{equation}
whereas the observed gap \cite{modesti,odobescu} is obviously hard-like. 

The origin of this discrepancy is related to very low conductivity value of the studied surface, $\sigma\ll e^2/h$. This leads to appearance of the zero-bias anomaly due to a finite time required for tunneling electron to leave the tunneling region, so that the tunneling DOS ({\it dI/dV} spectra)  does not correspond any more to the intrinsic DOS of a studied system, and therefore cannot be considered as its local DOS. For a diffusive conduction it leads to the logarithmic suppression of TDOS, $dI/dV\propto \ln |V|$~\cite{levitov}.  
To describe the TDOS of the Si(111)-$7\times7$ surface demonstrating the variable-range hopping conduction we employ the theory of the dynamical Coulomb blockade \cite{devoret, joyez, brun}. The theory accounts for the impedance of the equivalent electrical circuit in the tunneling current calculations. The respective circuit consisting of tunneling junction in series with an arbitrary external impedance $Z_{ext}(\omega)$ is shown in  (Fig.~\ref{LDOSdiffT}(j)). In our experimental scheme $Z_{ext}(\omega)=1/(i\omega C + R^{-1})$, where $C$ and $R$ are equivalent capacity and resistance of the surface. The tunneling current is calculated for the total impedance seen from the tunneling junction side, which is a parallel combination of $Z_{ext}$ and $C_T$, and is given by 
$$
Z(\omega) = [i\omega C_T+Z_{ext}^{-1}(\omega)]^{-1} = [i\omega C_{\sum}+1/R],
$$
where $C_{\sum} = C +C_T$. 
The $(dI/dV)R_T$ characteristics evaluated analytically according to \cite{joyez} are plotted in Fig.~\ref{LDOSdiffT} with green dotted line. As $R = 10^{7}-10^{13} \ \Omega$ is known from our conductivity measurements, so $C_\Sigma$ is \emph{the only fitting parameter} which is required to describe the entire data set shown in Fig.~\ref{LDOSdiffT}. The tunneling capacity was estimated to be not higher than $C_T \approx 0.3\cdot 10^{-19}$~F and therefore can be neglected. The best fit of our data was obtained for $C  = 1.8 \cdot 10^{-18}$~F. Then the localization region $a$ of tunneled electron could be estimated as  $a\sim C/8\pi\kappa\varepsilon_0\simeq 1.3 \ nm$. The value perfectly agrees with the localization length obtained from the transport measurements.

As seen from Fig.~\ref{LDOSdiffT} the dynamical Coulomb blockade theory describes  the experimental data inside the gap region fairly well, but it predicts an extra density of states above the gap edges. This disagreement is eliminated when the presence of the soft gap is taken into account  (Fig.~\ref{gapDiagr}(a)) in addition to the Coulomb blockade effect. In our case, the capacitance $C$ determines the barrier $e/2C$ for tunneling of the next electron. This modifies the Coulomb  gap (\ref{eq:DOS})  in TDOS spectra in a way which is presented in Fig. 2 of Ref~\cite{devoret} at $R/R_h = \infty$ (see Fig.~\ref{gapDiagr}(b)). 

\begin{figure}
\resizebox{59.5mm}{!}{\includegraphics{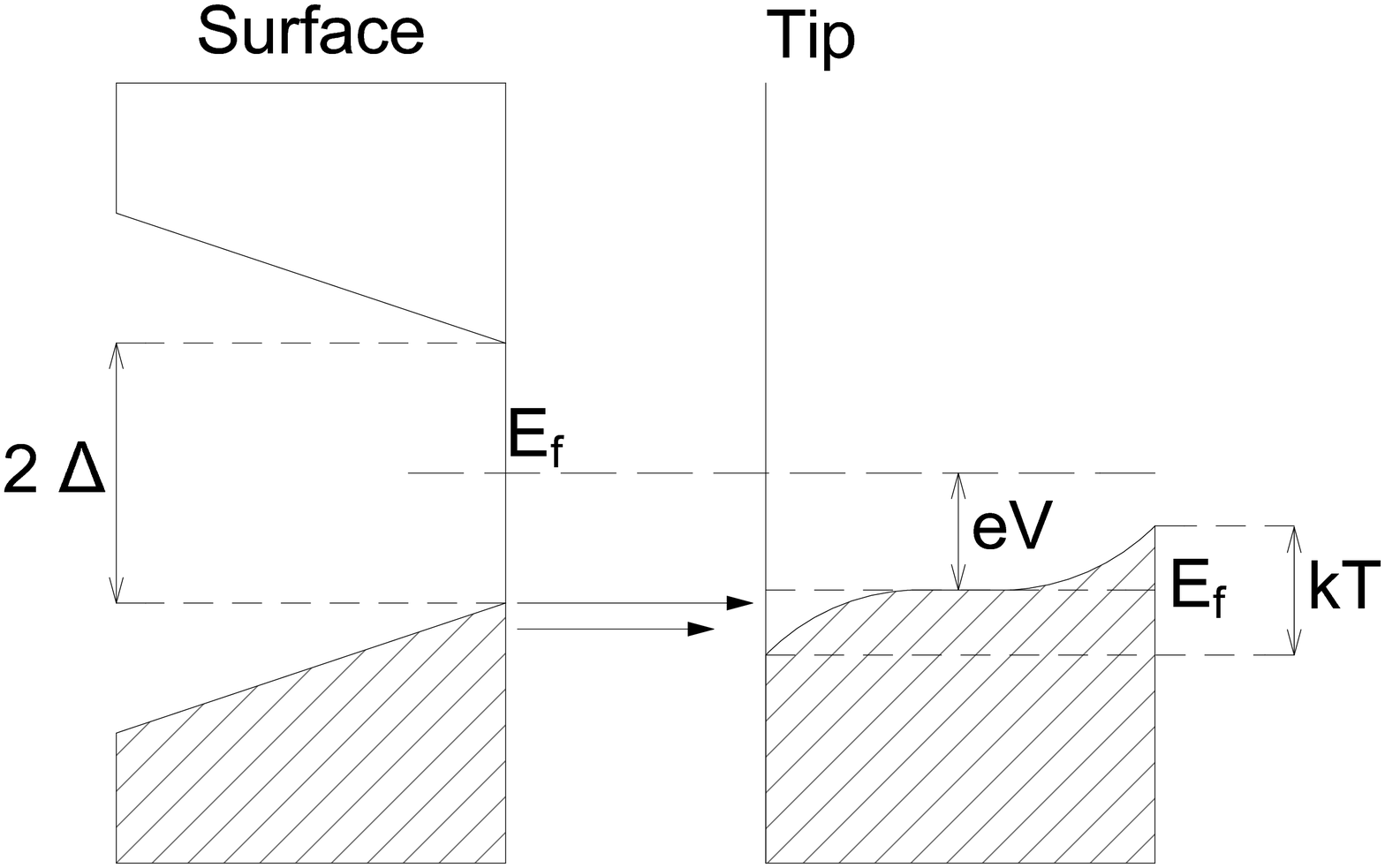}} 
\caption{Schematic energy diagram of the tunneling DOS of the Si(111)-$7\times 7$ surface and metallic STM tip. When $kT < 2 \Delta$ all states in the surface are occupied, when in the tip there is $kT$ spreading near $E_f$.}
\label{enDiakT}
\end{figure}

\begin{figure}
\resizebox{80mm}{!}{\includegraphics{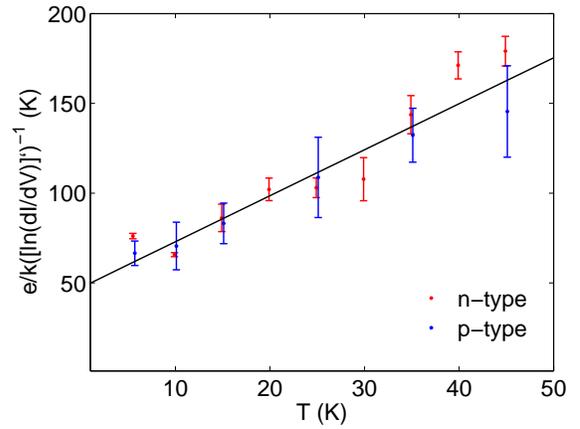}} 
\caption{The plot of incline of experimental $dI/dV$ data presented in semilogarithmic coordinates (Fig.~\ref{LDOSdiffT}) estimated at gap edge regions $V_{-} = -20 \pm 10$ mV and $V_{+} = 20 \pm 10$ mV. The derived proportionality $\alpha \sim 2.5$}
\label{inclines}
\end{figure}   

According this proposal we calculate the $(dI/dV)R_T$ in a simple case of tunneling between the tip and the surface with the TDOS shown in Fig.~\ref{enDiakT}. The tunneling current is given by
\begin{eqnarray}
\lefteqn{ I=D \int \limits^{+\infty}_{-\infty}g_s(E)g_t(E + eV)\left(\frac{1}{1 + e^{\frac{E}{\alpha kT}}}\right)\times } \nonumber
\\ & & \times\left(1 - \frac{1}{1 + e^{\frac{E + eV}{\alpha kT}}}\right)dE - D \int \limits^{+\infty}_{-\infty}g_s(E)g_t(E + eV)\times   \nonumber
\\ & &\times\left(1 - \frac{1}{1 + e^{\frac{E}{\alpha kT}}}\right)\left(\frac{1}{1 + e^{\frac{E + eV}{\alpha kT}}}\right)dE 
\label{eq:Itun}
\end{eqnarray}

where $g_t$ is the density of states in the metallic tip, $g_s$ is the surface density of states of the considered system (Fig.~\ref{enDiakT}), $D$ is the transmission coefficient  and $\alpha$ is the phenomenological parameter. The $\alpha$ could be found from the derivative $\ln (dI/dV)'\equiv d\ln (d I/dV)/dV$, which reflects the sensitivity of \textit{I-V} curves to thermal fluctuations.  If $kT  \ll  2\Delta$ and bias voltage $eV \lessapprox \Delta$, the tunneling occurs from surface states below the energy gap into the small tail $kT$ wide of unfilled states in the metallic tip (Fig.~\ref{enDiakT}). This smears out the gap edges and decreases the slope of the $dI/dV$ curves close to the energy gap edges region. The dependence of the slope $dI/dV$ on the temperature is
\begin{equation}
\frac{e}{k}\frac{1}{\left[\ln(\frac{dI}{dV})\right]'} \approx \alpha T.
\label{kT}
\end{equation}
The inverted slopes of the data of Fig.~\ref{LDOSdiffT} averaged over intervals $V_{-} = -20 \pm 10$ mV and $V_{+} = 20 \pm 10$ mV are presented in Fig.~\ref{inclines}. As seen, the experimental slope $\alpha$ is $2.5$ times larger than the expected value $\alpha \approx 1$. Taking it into account, we tried to fit the entire set of our STS data by using Eq. \ref{eq:Itun} with $\alpha\neq 1$. The best fit is obtained with experimentally determined values $2\Delta = 40$~meV and $\alpha=2.5$ providing quite reasonable approximation of the data in the entire voltage region (black dotted line in Fig.~\ref{LDOSdiffT}), whereas the Coulomb blockade model fits the data in the gap region only. That points out that the gap is actually not hard. Indeed, the low-voltage conductance in the Coulomb blockade case obeys the power law, $I\propto V^\beta$ \cite{devoret}, which can be hardly distinguished from the exponent if $\beta\gg 1$ (in our case $\beta\approx 4$ at 10~K). So our model (Fig.~\ref{gapDiagr}(b)) may be used as a reasonable approximation of the experimental results not only in the gap region but also at $V>\Delta$. Further study, both experimental and theoretical, required to develop a self-consistent model which describes tunneling into low-conducting surface with hopping conduction.

\section{Conclusion}
In conclusion, it was found that the temperature variation of the conductivity of the Si(111)-$7\times7$ surface follows the Efros-Shklovskii law $\ln(\sigma) \propto T^{-1/2}$  at least in $10-100$~K temperature range. The experimentally obtained localization length of the electron ($\xi=1.3$ nm) agrees with the one suggested earlier. 
The discrepancy between the tunneling differential conductance of the Si(111)-$7\times7$ surface and the intrinsic surface density of states is analysed. We propose, that the shape of the energy gap at Fermi level observed in the tunneling spectra is determined by both the Coulomb correlation gap and the local environmental Coulomb blockade effect. Two different phenomenological models describing the observed TDOS and its temperature evolution are presented. We consider that the observed difference between intrinsic DOS and TDOS obtained from tunneling spectroscopy is a general phenomena in a wide class of low-conducting surfaces.

\begin{acknowledgments}
We are grateful to D. Rodichev for stimulative discussion. The work was supported by the Russian Foundation for Basic Research  and the program of Department of Physical Sciences of RAS. 
\end{acknowledgments}

\end{document}